\documentclass[08pt, twocolumn,preprintnumbers,amsmath,amssymb,nofootinbib,superscriptaddress]{revtex4}
\usepackage{amsmath,amssymb,graphicx}
\usepackage{mathtools}
\usepackage{float}
\usepackage{latexsym} 
\usepackage{amsmath,amsthm}
\usepackage{ifpdf}
\usepackage{epstopdf}
\usepackage{dcolumn}% Align table columns on decimal point
\usepackage{bm}% bold math
\usepackage{braket}
\usepackage{graphicx}
\newcommand{\ha}{\hat{a}}%
\newcommand{\had}{\hat{a}^\dag}%
\newcommand{\barr}{\begin{eqnarray}}
\newcommand{\earr}{\end{eqnarray}}
\begin{document}
\def \beq{\begin{equation}}
\def \eeq{\end{equation}}
\def \bea{\begin{eqnarray}}
\def \eea{\end{eqnarray}}
\def \bes{\begin{split}}
\def \ees{\end{split}}
\def \besu{\begin{subequations}}
\def \esu{\end{subequations}}
\def \bea{\begin{align}}
\def \eal{\end{align}}
\newcommand{\fieldE}[2]{E(\underline{#1},#2)}
\newcommand{\fieldA}[2]{A(\bold{#1},#2)}
\newcommand{\fieldX}[3]{\psi_{#1,#2}^{(#3)}(R,\zeta)}
\newcommand{\fieldXa}[3]{\psi_{#1,#2}^{(#3)}(R,\zeta_1)}
\newcommand{\fieldXb}[3]{\psi_{#1,#2}^{(#3)}(R,\zeta_2)}
\newcommand{\creatX}[3]{\hat{a}_{#1#2}^{\dagger}(#3)}
\newcommand{\annX}[3]{\hat{a}_{#1#2}(#3)}
\newcommand{\creatXb}[3]{\hat{b}_{#1 #2}^{\dagger}(#3)}
\newcommand{\annXb}[3]{\hat{b}_{#1 #2}(#3)}
\newcommand{\annA}[2]{\hat{A}(\bold{#1},#2)}
\newcommand{\creatA}[2]{\hat{A}^{\dagger}(\bold{#1},#2)}

\title{Squeezing of  X waves with orbital angular momentum}
\author{Marco Ornigotti}
\affiliation{Institute of Applied Physics, Friedrich-Schiller University, Jena, Max-Wien Platz 1, 07743 Jena, Germany}
\author{Leone Di Mauro Villari}
\affiliation{University of Rome La Sapienza, Department of Physics, Piazzale Aldo Moro 5 00185, Rome, Italy}
\affiliation{Institute for Complex Systems, National Research Council, (ISC-CNR), Via dei Taurini 19, 00185, Rome (IT) }
\author{Alexander Szameit}
\affiliation{Institute of Applied Physics, Friedrich-Schiller University, Jena, Max-Wien Platz 1, 07743 Jena, Germany}
\author{Claudio Conti}
\affiliation{University of Rome La Sapienza, Department of Physics, Piazzale Aldo Moro 5 00185, Rome, Italy}
\affiliation{Institute for Complex Systems, National Research Council, (ISC-CNR), Via dei Taurini 19, 00185, Rome (IT) }
\begin{abstract}
Multi-level quantum protocols may potentially supersede standard quantum optical polarization-encoded protocols in terms of amount of information transmission and security.
However, for free space telecomunications, we do not have tools for limiting loss due to diffraction and perturbations, as for example turbulence in air.
Here we study propagation invariant quantum X-waves with angular momentum; this representation expresses the
electromagnetic field as a quantum gas of weakly interacting bosons. The resulting spatio-temporal quantized light pulses are not subject to diffraction and dispersion, 
and are intrinsically resilient to disturbances in propagation. We show that spontaneous down-conversion generates squeezed X-waves useful for quantum protocols.
Surprisingly the orbital angural momentum affects the squeezing angle, and we predict the existence of a characteristic axicon aperture for maximal squeezing.
There results may boost the applications in free space of quantum optical transmission and multi-level quantum protocols, and may also be relevant for novel
kinds of interferometers, as satellite-based gravitational wave detectors.
\end{abstract}
\maketitle
Standard protocols of quantum communication encode information into the polarization degrees of freedom of  photons \cite{BB,Ek}. As a result, only one bit of information can be imprinted onto each photon. In recent years, there has been great interest in the development of  a free-space system for quantum communications  based on the use of modes that carry orbital angular momentum (OAM) \cite{B,MM,MR,HR,GJ,MK,CM}.  When using OAM  there is no limit to the number of bits of information that can be carried by a single photon, as the OAM states span an infinite-dimensional space. 
Correspondingly, the rate of information increases drastically. In addition, the security of the considered  protocol  is increased by a multi-level basis  \cite{B}. A key problem related with multimode quantum communications is the diffraction and dispersion of the wave-packet. Diffraction and dispersion create an inhomogeneous transmission loss for different spatial frequencies that results in mixing of spatial modes\cite{MM}. Moreover it has been shown that OAM states are strongly affected by perturbations. A great deal of work has been done in studying the effect of atmospheric turbulence in free space communication \cite{MR,GA,TB,AN,KF}.  
A promising solution is the use of non-diffracting or localized waves such as X-waves that are naturally resilient against perturbation \cite{HZ}. 

Localized waves, i.e., linear solutions of Maxwell's equations that propagate without diffracting in both space and time, have been the subject of extensive research in the last years \cite{HZ}. In particular X-waves, firstly introduced in acoustics in 1992 by Lu and Greenleaf \cite{LG}, have been studied in different areas of physics \cite{CT,LH}. Despite the great amount of literature concerning X-waves, however, the investigations of their quantum properties are very few and they are limited to the case of traditional X-waves, without OAM  \cite{CC,CiC}. Very recently X-wawes carrying OAM have been proposed \cite{OC} and they can constitute a new possible platform for free space quantum communication. 

In this Letter, we present a quantum theory of X-waves carrying OAM based on the quantization of the motion of an optical pulse propagating in a normally dispersive medium. 
In particular, we study the case of spontaneous parametric down conversion (SPDC) in a quadratic medium, with particular attention to the effect of the OAM carried by quantum X-waves on the squeezing properties of the down converted stated generated by the nonlinear process. We find that  squeezing  is strongly affected by OAM; changing the parity of the OAM rotate the squeezing angle. This effect has a direct experimental signature and may be employed for novel quantum protocols.

 We start our analysis considering an electromagnetic field propagating in a medium with refractive index $n=n(\omega)$. Under the paraxial and slowly varying envelope approximation (SVEA), the field envelope $\fieldA{r}{t}$ satisfies the following  equation 
 \beq
 \label{eq2}
i\frac{\partial A}{\partial t}+i\omega'\frac{\partial A}{\partial z}-\frac{\omega''}{2}\frac{\partial^2A}{\partial z^2}+\frac{\omega'}{2k}\nabla_{\perp}^2A=0.
\eeq
We use  (\ref{eq2})  to study the propagation of an electromagnetic field in a dispersive medium characterized by a refractive index $n$, first order dispersion $\omega'=d\omega/dk$ and second order dispersion $\omega''=d^2\omega/dk^2$ \cite{KW}. For a field propagating in vacuum one has $\omega'=c$ and $\omega''=c^2/\omega$. The general solution of equation (\ref{eq2}) can be written as a polychromatic superposition of Bessel beams as follows  \cite{OC}
\begin{widetext}
\beq\label{eq3}
\fieldA{r}{t}=\sum_m\,d_m\,e^{im\theta}\,\int\,dk_z\,\int_0^{\infty}\,dk_{\perp}\,k_{\perp}\,S(k_{\perp},k_z)\,J_m(k_{\perp}R)\,e^{i(k_zZ-\Omega t)},
\eeq
 \end{widetext}
where  $Z=z-\omega't$ and $\Omega=-\omega''k_z^2/2+\omega'k_{\perp}^2/2k$. In the above equation  the cylindrical coordinates $\{R,\theta,Z\}$ and $\{k_\perp,\varphi,k_z\}$ have been used. This integral furnishes the field at instant $t$, given its spectrum $S(k_\perp,k_z)$ at $t=0$, with transversal and longitudinal wave-number $k_{\perp}$ and $k_z$,  respectively. If we introduce the change of variables $\{k_{\perp},k_z\}\rightarrow\{\alpha,v\}$ such that  $k_{\perp}=\sqrt{\omega''k/\omega'}\alpha$ and $k_z=\alpha-v\omega''$, with $v=c/\cos\vartheta$, being $\vartheta$ the Bessel cone angle, after some manipulation Eq. \eqref{eq3} can be rewritten  as a superposition of OAM-carrying X waves as follows:

%We introduce the following change of variables
%\beq
%\left\{\begin{array}{ll}
%k_{\perp}=\alpha\sqrt{\frac{\omega''k}{\omega'}}\\
%k_z=\alpha-\frac{v}{\omega''}
%\end{array}\right.
%\eeq
%where $\alpha$ and $v$ are two new variables which represent a normalized transverse $k$-vector and the X-wave velocity  defined by
%\beq \label{vel}
%v=\frac{c}{\cos \vartheta}
%\eeq	
%being $\vartheta$ is the axicon angle. After some manipulations the field $\fieldA{r}{t}$ can be written as a superposition of OAM-carrying X waves as follows 

\beq\label{eq4}
\fieldA{r}{t}=\sum_{m,p}\,\int\,dv\,C_{m,p}(v)e^{-\frac{iv^2t}{2\omega''}}\fieldX{m}{p}{v}
\eeq
where $\psi_{m,p}^{(v)}(R,Z-vt)$ is  the OAM-carrying X wave of order $p$ and velocity $v$ \cite{LL}
\begin{widetext}
\beq\label{eq5}
\fieldX{m}{p}{v}\equiv\int_0^{\infty}\,d\alpha\,\sqrt{\frac{k}{\pi^2\omega'(1+p)}}(\alpha\Delta)L_p^{(1)}(2\alpha\Delta)e^{-\alpha\Delta}\,J_m\left(\sqrt{\frac{\omega''k}{\omega'}}\alpha R\right)\,e^{i\left(\alpha-\frac{v}{\omega''}\right)\zeta}\,e^{im\theta}.
\eeq
\end{widetext}
Here, $L_p(x)$ are the generalized Laguerre polynomials of the first kind, $\zeta=Z-vt$ is the co-moving reference frame associated to the X-wave \cite{HZ} and   $\Delta$ is a reference length  related to the spatial extension of the beam. Following the orthogonality relation 
\beq
\langle\fieldX{l}{q}{u}|\fieldX{m}{p}{v}\rangle=\delta_{m,l}\delta_{p,q}\delta(u-v)
\eeq
we find that  OAM-carrying X waves have an infinite norm, like the plane waves typically adopted for field quantization. To quantize the field given by Eq. (\ref{eq4}) we employ the standard technique of expressing the total energy of the field as a collection of harmonic oscillators  \cite{CC,MW}. From Eq. \eqref{eq3} we find for the total energy carried by  $\fieldA{r}{t}$
\beq \label{eq6}
\mathcal{E}=\int\,d^3r\,|\fieldA{r}{t}|^2=\sum_{p,m}\,\int\,dv\,|\,C_{m,p}(v,t)\,|^2
\eeq
with $C_{m,p}(v,t)=C_{m,p}(v)e^{-\frac{iv^2}{2\omega''}t} $. As it can be seen the above equation  can be interpreted as a collection of harmonic oscillators with complex amplitude $C_{mp}(v)$ and frequency $\omega_m(v)=\frac{v^2}{2\omega''}$. Without loss of generality we perform  the quantization of the the fundamental X-wave  ($p=0$).  The generalization to $p \neq 0$ is straightforward. We  introduce  a pair of real canonical variables $Q_m(v,t)$ and $P_m(v,t)$ defined by
\beq
C_{m}(v,t)=\frac{1}{\sqrt{2}}\left[\omega_{m}(v,t)Q_{m,p}(v)+iP_{m}(v,t)\right],
\eeq
where $Q_m(v,t)$ and $P_m(v,t)$ oscillate sinusoidally in time at a frequency $\omega_m(v)$. We then obtain
\beq \label{eq5}
H =\frac{1}{2} \sum_{m}\,\int\,dv \left[P^2_{m}(v,t)+\omega_{m}^2(v)Q^2_{m}(v,t)\right].
\eeq
The total energy of the field can be, therefore, expressed as  an integral sum of harmonic oscillators characterized by the frequency $\omega_{m,p}(v)$, and $Q_{m,p}(v)$ and $P_{m,p}(v)$ play the role of position and momentum of the field, respectively. We promote these quantities to operators and introduce the creation and annihilation operators in the usual  way as
\besu
\begin{align}
\hat{Q}_{m}(v,t)&=\sqrt{\frac{\hbar}{2\omega_{m}(v)}}\left[\creatX{m}{}{v,t}+\annX{m}{}{v,t}\right],\\
\hat{P}_{m}(v,t)&=i\sqrt{\frac{\hbar\omega_{m}(v)}{2}}\left[\creatX{m}{}{v,t}-\annX{m}{}{v,t}\right],
\end{align}
\esu
 where $\annX{m}{}{v,t}=e^{i\omega_m(v) t }\annX{m}{}{v}$ and  the standard canonical bosonic commutation relations are understood \cite{MW}. Using the relations above, we obtain the Hamilton operator for the field from Eq. (\ref{eq5}) 
\beq
\hat H =  \hbar \sum_m \int dv\, \omega_m(v)\Bigl[ \creatX{m}{}{v}\annX{m}{}{v} + \frac{1}{2}\Bigl].
\eeq
 Hereafter we  drop the zero point energy as a standard renormalization procedure \cite{MS} . The above expression for the Hamilton operator describes the dynamics of an electromagnetic field expressed as a continuous superposition of harmonic oscillators. Each oscillator is characterised by a frequency $\omega_m(v)$ and it is associated to a travelling mode represented by Eq. \eqref{eq5}. Moreover, each travelling mode is parametrised with its velocity (i.e., axion angle) $v$.
 %
%each one characterized by the frequency $\omega_m(v)$ and associated to a travelling mode represented by Eq. (6) snd parametrized by its velocity (i.e., axicon angle) $v$. 
 % 
  A closer inspection to the above equation reveals a very intriguing fact, namely that the quantum dynamics of the field solution of Eq. (1) can be represented by a quantum gas of weakly interacting bosons with velocity $v$ and mass $M=\hbar/\omega''$. This is the first result of our Letter.
  
 We now substitute the expression of $C_m(v)$ in terms of $\creatX{m}{}{v}$ and $\annX{m}{}{v}$ into Eq. (\ref{eq4}) to obtain the field operator
 \beq\label{fieldOp}
\annA{r}{t}=\sum_{m}\,\int\,dv\,e^{-i \omega_m(v) t}\sqrt{\hbar\omega_{m}(v)}\psi_{m}^{(v)}(\bold r,\zeta)\annX{m}{}{v}.
\eeq
The above expression for the field operator can be intuitively understood as the result of the quantisation of the electromagnetic field in a cavity, where the normal modes of the cavity are represented by OAM-carrying X waves. We remark that this quantization approach is rigorous, and similar results can be obtained by using a standard Lagrangian approach, as we will report elsewhere.

As an example of application of the formalism developed above, we now consider the case of phase-matched, collinear spontaneous parametric down conversion (SPDC). In particular, we assume that a beam of frequency $\omega$ impinges upon a dielectric crystal with second order nonlinearity and we call it pump beam. As a result of the nonlinear interaction, a photon from the pump beam can be annihilated to create two new photons having lower frequencies $\omega_1$ and $\omega_2$ with 
$\omega=\omega_1+\omega_2$ \cite{MW}. Moreover, we assume that the non depleted pump approximation holds and that the pump beam can be therefore represented by a bright coherent state and treated like a classical beam. This allows us to consider its action on the Hamiltonian of the system as only a constant term, which can be therefore incorporated into the nonlinear coefficient $\chi$ associated to the process itself. Under these assumptions, the Hamiltonian describing such a system is then given by 
\beq
\begin{split}
\hat{H}=&\sum_{m}\,\int\,dv\,\hbar\omega_{m}(v)[\creatX{m}{}{v}\annX{m}{}{v}+\\ & +\creatXb{m}{}{v}\annXb{m}{}{v}]+\hat{H}_I(t),
\end{split}
\eeq
where the interaction Hamiltonian $\hat H_I(t)$ is obtained by quantizing its classical counterpart \cite{Boyd} 
\beq
\mathcal E_I = \chi \braket{A_1|A_2^*} + \text{c.c.}
\eeq
Here, the expression $c.c$ denotes the complex conjugation,  $\chi$ is  proportional to the second order nonlinearity $\chi^{(2)}$ and $\annX{m}{}{v}$ and $\annXb{m}{}{v}$ are the annihilation operator related to the quantized fields $\hat A_1(\bold r,t)$, $\hat A_{2}(\bold r, t)$ respectively. 
We assume that the two X-waves travel with the  same velocity $v$, such that the the two Bessel angles satisfy $\vartheta_2 = \vartheta_1 + 2n\pi$. After  lengthy but straightforward calculation, we can write the quantized  interaction Hamiltonian  as follows
\beq\label{intHam}
\hat H_I(t) = \hbar \sum_{m} \int dv \chi_{m}(2v) \omega_m(v) \hat a^\dag_{m}(v,t)b^\dag_{-m}(v,t) + \text{h.c.}
\eeq
In the expression above, $\text{h.c.}$ denotes the hermitian conjugate, while $\chi_m(x)$ is the interaction function, whose explicit expression reads
\beq
\chi_m(x) = (-1)^m4\pi^2 \chi xe^{-2x}.
\eeq
Moreover, in Eq. \eqref{intHam} we used $\annX{m}{}{v,t}=e^{iF(v) t}\annX{m}{}{v}$, with
\beq
F(v)= \frac{v^2}{\omega''}+\frac{v(1-\rho )(\omega_1'-\omega_2')}{\omega''(1+\rho)},
\eeq
where  $\rho=\sqrt{k_1\omega_1'/k_2\omega_2'}$, $\omega_{1,2}'=d\omega_{1,2}/dk$ and $k_{1,2}=\omega_{1,2}n_{1,2}/c$. 

To determine the electromagnetic field after the evolution driven by the interaction Hamiltonian $\hat{H}_I$, we employ  perturbation theory. Writing the total (time-dependent) Hamiltonian as
\beq
\hat{H}(t)=\hat{H}_0(t)+\lambda\hat{H}_I(t),
\eeq
then, using the Schwinger-Dyson expansion truncated at the first order, we get the following result for the state of the system \cite{MS}:
\beq
\ket{\psi^{(1)}(t)}=-\frac{i}{\hbar}\int_0^t\,d\tau\,\hat{H}_I(\tau)\ket{0}
\eeq
where $\ket{\psi(0)}=\ket{0}$ has been assumed. If we now introduce the quantities 
\barr
K(v)&=&\frac{v(1-\rho) (\omega_1'-\omega_2')}{2\omega''(1+\rho)}\\
G(v,t)&=&-\frac{2i}{F(v)}\sin\left[\frac{F(v)t}{2}\right],
\earr
and  we define the function
\beq
\mathcal{G}_{m}(v,t)=\sqrt{\omega_{m}(v)\omega_{-m}(v)}G(v,t)e^{iK(v)t}\chi_{m}(2v),
\eeq
after some algebra the final expression for the state after the interaction is 
\beq \label{eq7}
\ket{\psi^{(1)}(t)}=\sum_{m}\,\int dv\,\mathcal{G}_{m}(v,t)|m,v;-m,v\rangle
\eeq
where $|m,v;-m,v\rangle\equiv\creatX{m}{}{v}\creatXb{-m}{}{v}\ket{0}$.
The state given by Eq. (\ref{eq7}) represents a superposition of two particles, corresponding to the two modes $\omega_1$ and $\omega_2$ traveling with the same velocity $v$. Notice, moreover, that due to angular momentum conservation, the photon pairs generated by SPDC are constrained to possess the same amount of OAM but with opposite sign. This is possible since we have made no particular assumption about the OAM content of the pump beam. In the general case, in fact, the conservation of OAM implies that $m_s+m_i=m_p$, where the subscript $s$,$i$ and $p$ stand for signal, idler and pump, respectively.

 We can now study the squeezing effect in the case of degenerate down conversion, corresponding to $\omega_1=\omega_2=\omega/2$. The quantized field operator associated to the mode $\omega/2$ is then given as follows \cite{DH}:
\beq
\hat A(\bold r,t) = \sum_{m,p} \int dv \, \sqrt{\hbar \omega_{m,p}(v)} \psi^{(v)}(R,\zeta) \hat a_{m,p}(v,t).
\eeq\
In this case the Hamiltonian of the system is  the same as the one presented in Eq. \eqref{intHam} with $b_{m}(v)=a_{m}(v)$ and $\rho=1$, since we are considering the degenerate down conversion in which $\omega_1'=\omega_2'$ and $|k_1|=|k_2|$.
In the interaction picture we consider only the time evolution controlled by $H_I$ \cite{MS}. Thus the two  equations of motion for $\ha_{m}(v,t)$ and $\ha_{-m}(v,t)$ are  then
\beq
\begin{aligned}
\frac{\displaystyle d}{\displaystyle dt} \ha_{m}(v,t)&=\omega_{m}(v) [\chi_{m}(2v)+ \chi_{-m}(2v)]\had_{-m}(v,t),\\
\frac{\displaystyle d}{\displaystyle dt} \ha_{-m}(v,t)&=\omega_{m}(v) [\chi_{m}(2v)+ \chi_{-m}(2v)]\had_{m}(v,t).
\end{aligned}
\eeq
A general solution of these equations can be written as follows \cite{MW}:

\besu \label{sol}
\begin{align}
\hat a_{m}(v,t)&= \mathcal{A}_m(v,t)\ha_{m}(v)+\mathcal{B}_m(v,t)\had_{m}(v),\\
\hat a_{-m}(v,t)&= \mathcal{A}_{-m}(v,t)\ha_{m}(v)+\mathcal{B}_{-m}(v,t)\had_{m}(v),
\end{align}
\esu
where $\mathcal{A}_m(v,t)=\cosh[\xi_m(v)t]$, $\mathcal{B}_m(v,t)=e^{i(\phi+m\pi)}\sinh[\xi_m(v)t]$ and the squeezing parameter $\xi_m(v)$ is given, for a fundamental X wave, as follows:
\beq \label{sqParam}
\xi_{m}(v) = (-1)^m \frac{\pi \chi c}{2\Delta } e^{-\frac{\tilde v}{\omega''}}\frac{\tilde v^3}{(\omega'')^3},
\eeq
being  $\tilde v = v \Delta $. Equation \eqref{sol} shows that the state after the SPDC is a squeezed state with a squeezing parameter $\xi_m(v)$ depending on both velocity and OAM (see Fig.\ref{fig} below). To further elaborate on that, we can introduce the quadrature operators: 
$
\hat X_{j}(v,t) =  \ha_{j}(v,t)+\had_{j}(v,t) \,\,\, \text{and} \,\,\,
\hat Y_{j}(v,t) = i[\had_{j}(v,t) - \ha_{j}(v,t)]
$,
where $j=\{-m,m\}$. For $\phi=0$ we have 
\besu
\begin{align}
\hat X_{j}(v,t)&=e^{ \xi_{j}(v)t} \hat X_{j}(v,0),\\
\hat Y_{j}(v,t)&=e^{- \xi_{j}(v)t} \hat Y_{j}(v,0),
\end{align}
\esu
\begin{figure}[H]
\begin{center}
\includegraphics[width=1 \columnwidth]{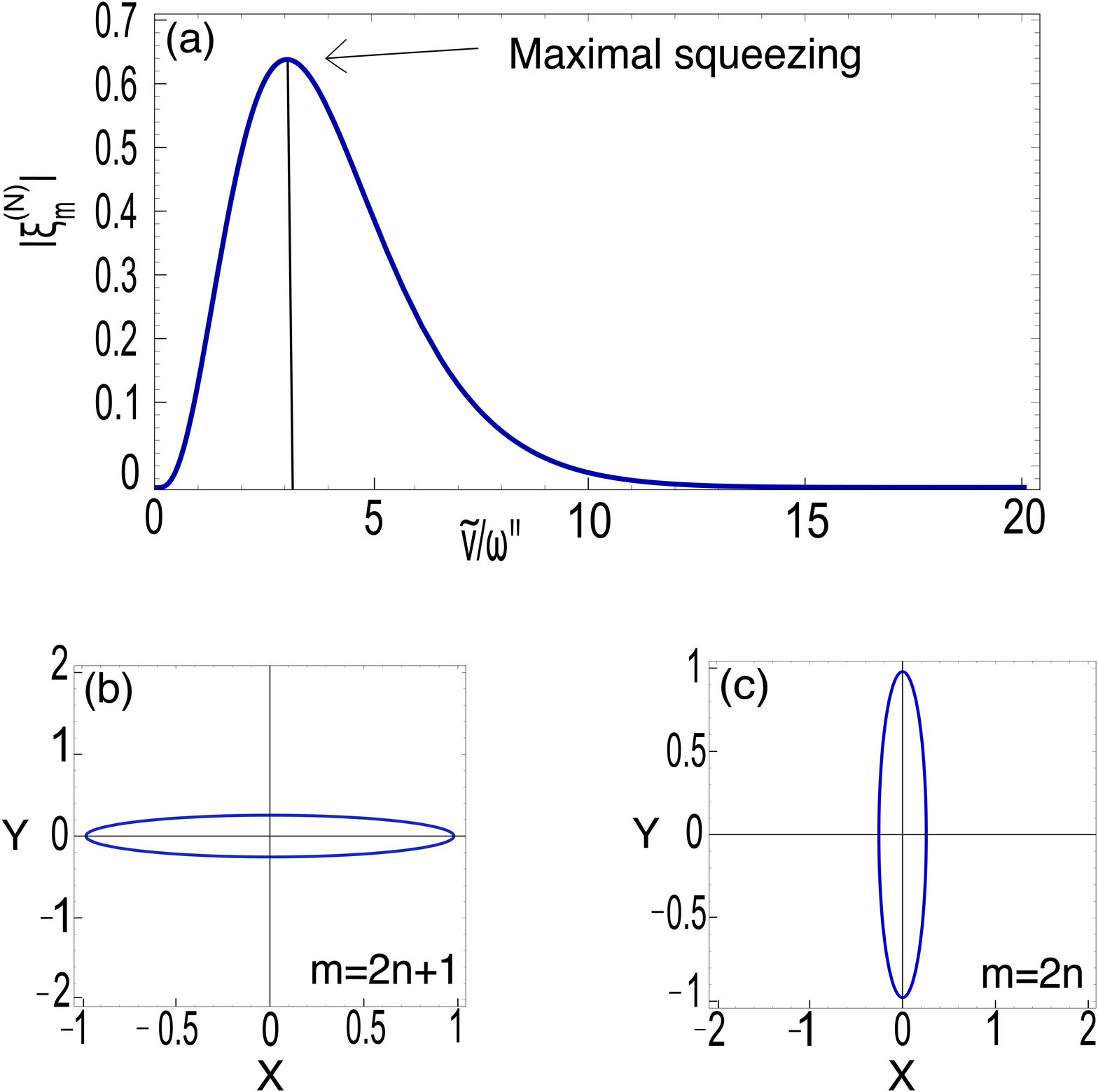}
\caption{(a) Plot of the of the normalized squeezing parameter modulus $|\xi^{(N)}_m|=\Delta|\xi_m|/(\pi\chi c)$  in function of the normalized velocity $\tilde v/\omega''$. (b),(c) Quadrature space representation of the squeezed down-converted state in the case of odd [panel (b)] and even [panel (c)] values of the OAM parameter $m$ with a normalized velocity $\tilde v/\omega''=3$ and fixing the time $t$ so that $\pi\chi ct/\Delta=1$.}
\label{fig}
\end{center}
\end{figure}
and the variance of such quadrature operators is then given by  $\Delta X_j(v,t)=e^{\xi_m(v)t}\Delta X_j(v,0)$ and $\Delta Y_j(v,t)= e^{-\xi_m(v)t}\Delta Y_j(v,0)$. This shows that the down conversion interaction Hamiltonian for OAM carrying X-waves acts like a two mode squeezing operator. Remarkably, we find that OAM changes the sign of the squeezing parameter $\xi_m(v)$ ,i.e., the squeezed quadrature changes depending on the parity of the angular momentum number $m$. In particular, if $m$ is an even number, $\xi_m(v) > 0$ and the squeezing occurs in the $Y$-quadrature. On the other hand, if $m$ is an odd number, $\xi_m(v) < 0$ and the $X$-quadrature will result squeezed as we can observe in figure \ref{fig} (b) and (c).This is the second result of our Letter. 

In addition, Eq. \eqref{sqParam} reveals a dependence of the squeezing parameter from the X wave velocity. Therefore, there exists an optimal value of the velocity 
 $v_{opt}=3\omegaÕÕ/\Delta$ that maximizes the amount of squeezing produced by the nonlinear process (Fig. \ref{fig}a). This corresponds to the optimal axion angle
$\cos \vartheta_{0}^{opt} = \Delta / 3\lambda$. If we, for example, assume a nondiffracting pulse with a duration of $\Delta t=8$ $fs$ and a carrier wavelength of $\lambda = 850$ $nm$, the optimal axicon angle that maximizes the squeezing is given by $\vartheta_0^{opt}\simeq 20^{\circ}$.  Using these values and assuming for the second order nonlinearity $ \chi^{(2)}\cong 10^{-12}\, \frac{m}{V}$ \cite{Boyd}, we can evaluate the maximal squeezing parameter to be $\xi_m \simeq 100 s^{-1}$.
 
We remark that the experimental generation of the proposed quantum states of light may be implemented by using a spiral phase-plate and a system of cylindrical lenses to control the OAM carried by the pump beam \cite{BC,AB}. The spiral phase-plate transform a $\text{TEM}_{00}$ mode in a spiral mode with fixed OAM \cite{BC}. The cylindrical lenses transform an input mode with a fixed OAM number $m$ (e.g., a Laguerre-Gauss mode) into one with number $-m$ \cite{AB}. In this way we can generate  two input beams, one with OAM per photon $\hbar m$ and one with OAM $-\hbar m$ per photon, that  are sent to the nonlinear crystal for SPDC. Another way to  realize X-waves carrying OAM is the use of metasurfaces to convert the spin angular momentum (SAM) in OAM \cite{BL,YL}. Suppose we have an input X-wave with $m=0$ and uniform circular polarization; after  the interaction with the metasurface  the output beam switches handedness with a SAM variation $\pm 2\hbar$ per photon. Since the total angular momentum must be conserved an OAM $m=\pm 2\hbar$ per photon is generated. The result for the field amplitude is a X-wave carrying OAM.  Alternatively the same result can be achieved by using total internal reflection in an isotropic medium \cite{BM}.

In conclusion, we have presented a quantized theory of optical pulses propagating in a normally dispersive medium as a collection of harmonic oscillators associated to to travelling modes represented by X waves carrying OAM. This allows us to describe the dynamics of the quantised field as  the ones of a one dimensional quantum gas of weakly interacting bosons with velocity $v=c/\cos\vartheta$ and mass $M=\hbar/\omega''$. Moreover, we have shown that it is possible to select the quadrature  squeezed state generated by SPDC [Figs. \ref{fig} (b) and (c)] and that there exists  an optimal velocity (i.e.,  axicon angle) that maximises the amount of squeezing generated. The presented theory provides the way to find the optimum angle for maximizing the squeezing effect. We believe that these results are helpful for future multilevel, free space, quantum communication protocols that are potentially free of diffraction and dispersion and not affected from external perturbations in particular from atmospheric turbulences.
Further applications include the use of the proposed diffraction-free OAM states in free space interferometric setups for high-sensitivity interferometers for gravitational wave detection. 

A.S. gratefully acknowledges financial support from the Deutsche
Forschungsgemeinschaft (grants SZ 276/7-1, SZ 276/9-1, BL 574/13-1, GRK 2101/1) and the German Ministry for Science and Education (grant 03Z1HN31);
C.C. gratefully acknowledges financial support from the Templeton foundation (grant number 58277).

\bibliography{Bibliografia}
\end{document}